\documentclass[journal,12pt,onecolumn,draftclsnofoot,]{IEEEtran}
\usepackage{booktabs,ragged2e}
\usepackage[flushleft]{threeparttable}
\usepackage{url}
\usepackage{ulem}
\usepackage{lipsum}
\usepackage{graphics,graphicx}
\ifCLASSOPTIONcompsoc
\usepackage[caption=false,font=normalsize,labelfon
t=sf,textfont=sf]{subfig}
\else
\usepackage[caption=false,font=footnotesize]{subfi
	g}
\fi

\usepackage{cite}

\usepackage[numbers,sort&compress]{natbib}	
\ifCLASSINFOpdf
\else
\fi
\usepackage{epsf,psfrag,amssymb,amsfonts,color}
\usepackage{amsmath}
\usepackage{amsthm}

\usepackage[mathscr]{eucal}
\usepackage{algorithm}
\usepackage{algorithmic}
\usepackage{dsfont}
\usepackage{bbm}
\usepackage{float}
\usepackage{array}
\usepackage{fixltx2e}
\usepackage{stfloats}
\usepackage{url}
\usepackage{cleveref}
\usepackage{siunitx}
\usepackage{multirow}
\begin{document}

%
\title{Optimal Eco-driving Control of Autonomous \\ and Electric Trucks in Adaptation to Highway Topography: Energy Minimization and \\ Battery Life Extension}
%
%
%

\author{Yongzhi~Zhang,~\IEEEmembership{Member,~IEEE,}
        Xiaobo~Qu,~\IEEEmembership{Senior Member,~IEEE,}
        and~Lang~Tong,~\IEEEmembership{Fellow,~IEEE}
\thanks{Y. Zhang is with the College of Mechanical and Vehicle Engineering, Chongqing University, Chongqing,
400044, China and is also with the Department of Architecture and Civil Engineering, Chalmers University of Technology, Gothenburg,
41296, Sweden (yzzhangbit@gmail.com).}
\thanks{X. Qu is with the Department of Architecture and Civil Engineering, Chalmers University of Technology, Gothenburg,
	41296, Sweden (\textit{corresponding author}: xiaobo@chalmers.se).}
\thanks{L. Tong is with the School of Electrical and Computer Engineering, Cornell University, Ithaca, New York, 14853, USA (lt35@cornell.edu)}}

%
%

\markboth{Eco-driving control, \today}%
{Shell \MakeLowercase{\textit{et al.}}: Bare Demo of IEEEtran.cls for IEEE Journals}
%



\maketitle

\begin{abstract}
This paper develops a model to plan energy-efficient speed trajectories of electric trucks in real time by taking into account the information of topography and traffic ahead of the vehicle. In this real time control model, a novel state-space model is first developed to capture vehicle speed, acceleration, and state of charge. An energy minimization problem is then formulated and solved by an alternating direction method of multipliers (ADMM) that exploits the structure of the problem. A model predictive control (MPC) framework is further employed to deal with topographic and traffic uncertainties in real-time. An empirical study is finally conducted on the performance of the proposed eco-driving algorithm and its impact on battery degradation. The simulation results show that the energy consumption by using the developed method is reduced by up to 5.05\%, and the battery life extended by more than 100\% compared to benchmarking solutions.
\end{abstract}

\begin{IEEEkeywords}
Autonomous and electric trucks, heavy duty truck, speed control, energy minimization, battery life extension, MPC, ADMM.
\end{IEEEkeywords}

%
\IEEEpeerreviewmaketitle

\section{Introduction}
%
%
%
%
\IEEEPARstart{H}{eavy}-duty trucks (HDTs) account for 70\% of all freight transport and 20\% of transportation-sector greenhouse gas (GHG) emissions in the United States \cite{quiros2017greenhouse}. Therefore, the decarbonisation of HDTs is essential for developing sustainable transportation and one technology that could deliver this is the battery electric (BE) HDT \cite{earl2018analysis,sen2017does}. The electrification of HDT has developed greatly in recent years, and according to McKinsey, it was reported that battery electric trucks could account for 15\% of global truck sales by 2030 \cite{tryggestad2017new}. However, the development of BE HDT is greatly limited to the present lithium-ion battery technology especially owing to the low energy density and short cycle life \cite{sripad2017performance}.

One way to address this technology bottleneck is to develop the next-generation battery with largely improved energy density and cycle life \cite{lopez2019designing}, which however generally requires decades and cannot meet the requirement of immediate usage. Another way is to reduce vehicle energy consumption and battery aging through intelligent energy management and speed control strategies, which is owed to the eco-driving control techniques. Indeed, many algorithms have been developed for the eco-driving of both passenger vehicles \cite{dib2014optimal,xu2018cooperative,zeng2018globally,malikopoulos2018optimal} and HDTs \cite{hellstrom2009look,guo2018fuel,held2018optimal,borek2019economic,turri2016cooperative}. In \cite{dib2014optimal} the eco-driving techniques were discussed and formulated as an optimal control problem that consisted of the minimization of the vehicle consumption over a time and distance horizon, and then a closed-form solution of the optimal trajectories was derived. Xu et al. \cite{xu2018cooperative} proposed a cooperative method of traffic signal control and vehicle speed optimization for connected automated vehicles, which optimized the traffic signal timing and vehicles’ speed trajectories at the same time. Zeng et al. \cite{zeng2018globally} proposed the optimal speed planning solution for a vehicle running on a given route with multiple stop signs, traffic lights, et al., while Malikopoulos et al. \cite{malikopoulos2018optimal} addressed the problem of controlling the speed of a number of automated vehicles before they entered a speed reduction zone on a freeway. Although involving different driving scenarios with traffic signals, speed limits, etc. \cite{dib2014optimal,xu2018cooperative,zeng2018globally,malikopoulos2018optimal}, for model simplification purposes the passenger vehicle modeling generally neglects the impacts of road topographies and aerodynamics, which however greatly affect the energy consumption of HDTs featured with a heavy and large body \cite{lammert2014effect}. Some literature developed battery aging-conscious energy management strategies for hybrid electric vehicles (HEVs) by optimally splitting the driving power \cite{tang2015energy,sockeel2018pareto,zhang2021bi,guo2021real}, but how to drive the pure electric vehicles (EVs) in an energy-efficient way with minimized battery aging still remains an open problem. In fact, due to the long-haul driving requirement, the shortcomings of batteries such as low energy density and short cycle life is greatly enlarged with the BE HDT. An eco-driving algorithm considering battery aging is thus especially required for the BE HDT optimal control.

The eco-driving control of HDTs are mainly focused on the internal combustion engine (ICE)-powered HDTs. Hellström et al. \cite{hellstrom2009look} developed a predictive cruise controller where the dynamic programming (DP) method was used to solve the optimal control problem numerically. In ref. \cite{hellstrom2009look}, a pre-processing algorithm was developed to downsize the search space of DP so that the algorithm complexity was reduced for real-time operation. Guo et al. \cite{guo2018fuel} investigated the problem of speed planning and tracking control of a platoon of trucks and presented a two-layered hierarchical framework for truck platoon coordination: a speed planning layer for en route speed profile calculation and a control layer for vehicle speed tracking. Held et al. \cite{held2018optimal} developed fuel efficient driving algorithms for applications with varying speed limits in urban driving, while Borek et al. \cite{borek2019economic} developed economic optimal control strategies with traffic involved and navigation at signalized intersections by using infrastructure-to-vehicular communication. The speed control of traditional HDT generally involves the optimization of gear selections (integer), when the DP method is often introduced \cite{hellstrom2009look,guo2018fuel,borek2019economic,turri2016cooperative}. Compared to traditional HDTs, the speed control of BE HDT generally involves non-integer optimization, which thus provides the chance for introducing more efficient optimization algorithms than the DP method. To the best knowledge of the authors, this is the first paper focused on the optimal eco-driving control of BE HDT, in consideration of both the energy consumption and battery aging.

In this work three major contributions are made to optimizing truck speed trajectory. First, a novel state-space model is constructed to capture the dependencies of vehicle speed, acceleration, and battery state-of-charge (SOC). Based on this model, the optimization problem for energy minimization is further formulated, with both the road topographies and surrounding traffics involved. Then, a model predictive control (MPC) approach based on an alternating direction method of multipliers (ADMM) is developed by taking into account topographical and traffic uncertainties, so that the efficient real-time speed control is realized. Finally, the effects of the speed control algorithm on battery degradation is systematically evaluated by introducing an EV-oriented (battery) aging model, and an empirical study is further conducted to validate the performance of the eco-driving strategy with and without surrounding traffics, respectively.

The rest of the paper is organized as follows. Section II describes a state-space model describing truck system dynamics. Section III defines the energy optimization problem and develops optimization and control algorithms. Section IV shows the truck energy consumption results based on the developed method, followed by Section V showing the battery aging evaluation results. Section VI concludes the paper.

\section{State-space equation modeling}
In this section, a state-space model connecting the vehicle speed, acceleration and energy consumption is constructed, where the vehicle speed and battery SOC are the model state and output, respectively. The modeling processes are presented as below in details.

\begin{figure}[b]
	\centering
	\includegraphics[width=3.5 in]{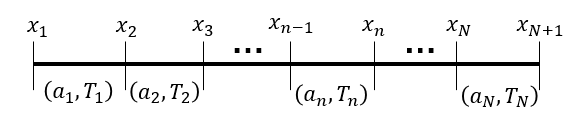}
	\caption{Schematic diagram.}
	\label{roadseg}
\end{figure}

Let the total trip be divided into $ N $ equally spaced segments of unit length, and segment $ i $ begins at $ t_i $. The schematic diagram is shown in Fig. \ref{roadseg} and the definition of segment variables is listed in Table \ref{segdef}.

\begin{table}[h]
	\begin{threeparttable}
		\caption{Segment variables definition for segment $ i $}
		\label{segdef}
		\setlength\tabcolsep{0pt}
		\begin{tabular*}{\columnwidth}{@{\extracolsep{\fill}}cll}
			\toprule
			Notation & Definition & Explanation \\
			\midrule
			$ T_i $ & The travel time trough segment $ i $ & - \\
			$ t_i $ & The starting time of segment $ i $ & $ t_1=0, t_{i+1}=t_i+T_i $ \\
			$ v_i(t) $ & \multicolumn{1}{p{7cm}}{The instantaneous EV velocity in segment $ i $} & $ t_i<t<t_{i+1} $ \\
			$ x_i $ & \multicolumn{1}{p{7cm}}{The initial velocity at the beginning of segment $ i $}  & \multicolumn{1}{p{7cm}}{$ x_i=v_i(t_i) $, it is the system state.} \\
			$ a_i $ & \multicolumn{1}{p{7cm}}{The EV acceleration in segment $ i $ } & \multicolumn{1}{p{7cm}}{This value is the system input and assumed to be constant in the short segment.} \\
			$ p_i(t) $ & \multicolumn{1}{p{7cm}}{The instantaneous power of the battery output in segment $ i $ } & \multicolumn{1}{p{7cm}}{Negative in charging and positive in discharging.} \\
			$ I_i(t) $ & \multicolumn{1}{p{7cm}}{The instantaneous current of the battery output in segment $ i $} & \multicolumn{1}{p{7cm}}{Negative in charging and positive in discharging.} \\
			$ y_i $ & \multicolumn{1}{p{7cm}}{The SOC change in segment $ i $} & \multicolumn{1}{p{7cm}}{This value is the system output.} \\
			\bottomrule
		\end{tabular*}
	\end{threeparttable}
\end{table}

The state-space model is defined as:
\begin{eqnarray}
	x_{i+1}=F_{i}(x_i,a_i), \; y_i=G_{i}(x_i,a_i),\nonumber
\end{eqnarray}
where $ F_{i}(\cdot,\cdot) $ is the state transition and $ G_{i}(\cdot,\cdot) $ is the output function.

\subsection{State transition function}
Note that the length of each road segment in Fig. \ref{roadseg} is considered as unit length ‘1’ for easy derivation purposes, but will be parameterized in the empirical studies {\color{red}(Section \ref{sec: case})} based on the practical conditions. In this case, $ x_i $, $ a_i $ and $ T_i $ are related by
\begin{eqnarray}\label{state1}
	\int_{0}^{T_i}(x_i+a_it)\,dt=1 \; \Rightarrow \; a_{i}T_{i}^2+2x_{i}T_{i}-2=0,
\end{eqnarray}
which gives
\begin{eqnarray}
	a_i=\frac{2}{T_{i}^2}-\frac{2x_i}{T_i}, \, T_i=\frac{-x_i+\sqrt{x_{i}^2+2a_i}}{a_i}.\nonumber
\end{eqnarray}
The state transition is then given by
\begin{eqnarray}\label{state}
	x_{i+1} = x_i+a_{i}T_{i}=\sqrt{x_{i}^2+2a_i} \; \Rightarrow \; F_{i}(x_i,a_i) = \sqrt{x_{i}^2+2a_i}.
\end{eqnarray}

\subsection{Output function}
\subsubsection{SOC change within each road segment}
SOC indicates the ratio of battery remaining capacity to nominal capacity, thus the SOC change within each road segment is defined as
\begin{eqnarray} \label{socch}
	y_i=G_{i}(x_i,a_i)=\frac{\int_{0}^{T_i}I_{i}(t)\,dt}{C}=\frac{\int_{0}^{T_i}p_{i}(t)\, dt}{UC},
\end{eqnarray}
where $ C $ represents the battery nominal capacity, and $ U $ represents the battery terminal voltage which is assumed constant within the commonly used SOC ranges \cite{berecibar2016state}.

\subsubsection{Longitudinal dynamics of vehicle}
The equation describing the longitudinal dynamics of vehicles is shown as:
\begin{eqnarray} \label{dynamics}
	F_{i}^{\rm track}(t)-F_{i}^{\rm air}(t)-F_{i}^{\rm res}=ma_i,
\end{eqnarray}
with
\begin{eqnarray}
	\begin{cases}
		F_{i}^{\rm air}(t)=\frac{1}{2}\rho_{\rm air}A_{\rm f}C_{\rm D}v_{i}^{2}(t), \nonumber \\
		F_{i}^{\rm res}=mgC_{\rm r}cos(\alpha_i)+mgsin(\alpha_i), \nonumber \\
	\end{cases}
\end{eqnarray}
where $ m $ is the vehicle mass, $ F_{i}^{\rm track}(t) $ is the track force generated by the electric motor, $ F_{i}^{\rm air}(t) $ is the air resistance with $ \rho_{\rm air} $ indicating the air mass density, $ A_{\rm f} $ indicating the vehicle frontal area and $ C_{\rm D} $ indicating the aerodynamic drag coefficient, and $ F_{i}^{\rm res} $ is the resistance consisting of the road frictional resistance and the gravitational resistance with $ C_{\rm r} $ indicating the rolling resistance factor and $ \alpha_i $ indicating the slope of segment $ i $. To simplify the derivation process, it is assumed that $ \beta^{\rm air}=\frac{1}{2}\rho_{\rm air}A_{\rm f}C_{\rm D} $ and $ F_{i}^{\rm res}=\beta_{i}^{(0)} $, where $ \beta^{\rm air} $ is independent of road segment conditions while $ \beta_{i}^{(0)} $ depends on the slope of each road segment.

\subsubsection{Output function modeling}
The track force $ F_{i}^{\rm track}(t) $ is produced by the battery power. When the battery discharges, $ p_{i}(t) > 0 $, the electric power is converted to positive track force according to
\begin{eqnarray}
	F_{i}^{\rm track}(t)=\beta^{+}p_{i}(t)/v_{i}(t), \nonumber
\end{eqnarray}
where $ \beta^{+} $ indicates the vehicle discharging efficiency.

When the battery is charged by the kinetic energy,  $ p_{i}(t) < 0 $, the motor produces negative track force as
\begin{eqnarray}
	F_{i}^{\rm track}(t)=\beta^{-}p_{i}(t)/v_{i}(t), \nonumber
\end{eqnarray}
where $ 1/\beta^{-} $ indicates the vehicle charging efficiency. Here, we assume that all the brakings can be accommodated through the electric brakes and are regenerative brakings; thus, the friction braking is not included. This assumption is consistent with the practice, where the use of friction brakes is minimized to improve energy efficiency and reduce wear \cite{amini2019cabin}.

Then the instantaneous battery power output is expressed as
\begin{eqnarray}
	p_{i}(t)=\frac{1}{\beta_{i}}v_{i}(t)F_{i}^{\rm track}(t), \, \beta_{i}=
	\begin{cases}
		\beta^+, \, F_{i}^{\rm track} \geq 0 \\
		\beta^-, \, {\rm otherwise}
	\end{cases}
\end{eqnarray}

where $ \beta_i $ includes the efficiencies of the battery charging/discharging, the AC/DC converter, the electric motor (EM), and the transmission gearbox. The transmission energy losses caused by the gearbox are small and the transmission energy efficiency is set to a constant close to 100\% \cite{turri2016cooperative}. The EM efficiency varies as the torque and rotational speed of the motor, which can also be considered a constant based on the motor efficiency map and the vehicle speed limit in this study. The detailed simplification process of the EM efficiencies is presented in Appendix \ref{app:EM}.

Note that the sign of track force can be verified from \eqref{dynamics}, thus whether $ \beta_{i} $ equals $ \beta^+ $ or $ \beta^- $ is determined by the vector of $ (x_i,a_i) $. That is, the track force can be positive, negative, changing from positive to negative or vice verse within each road segment. Theses four cases lead to different modeling processes of output function, which are presented in details in Appendix \ref{app:output}.

Generally, the output function can be expressed as
\begin{IEEEeqnarray}{rCl}
	G_i(x_i,a_i)=\gamma_{i}^{(0)}+\gamma_{i}^{(1)}a_{i}+\gamma_{i}^{(2)}x_{i}^{2}, \nonumber
\end{IEEEeqnarray}
where the expressions of vector $ \gamma_i(x_i,a_i) $ are concluded in Table \ref{gammaval} (Appendix \ref{app:output}).

\section{Formulation and solution of the energy minimization problem}
\subsection{Problem formulation of energy minimization}
The total energy consumed ($ E $) is given by the cumulative SOC change. In this case,
\begin{IEEEeqnarray}{rCl}
	E=\sum_{i=1}^{N+1}\left(\gamma_i^{(0)}+\gamma_i^{(1)}a_i + \gamma_i^{(2)}x_i^2 \right).
\end{IEEEeqnarray}

From \eqref{state1} and \eqref{state}, there is 
\begin{IEEEeqnarray}{rCl}
	\begin{array}{ccc}
		\begin{cases}
			x_{i+1}=x_i +a_i T_i \\
			a_i T_i^2+2x_i T_i-2=0 
		\end{cases} & \Rightarrow &
	    \begin{cases}
	    	a_i=\frac{x_{i+1}^2-x_i^2}{2} \nonumber\\
	    	\frac{1}{T_i}=\frac{x_{i+1}+x_i}{2}. \nonumber
	    \end{cases}
	\end{array}
\end{IEEEeqnarray}

By substituting $ a_i $ in the total energy expression, there is
\begin{IEEEeqnarray}{rCl}
	E=\eta_0(x)+\sum_{i=1}^{N+1}\eta_i(x)x_i^2,
\end{IEEEeqnarray}
where $ x=\left[x_1,x_2,\cdots,x_{N+1}\right]^\top $, and
\begin{IEEEeqnarray}{rCl}
	\begin{cases}
		\eta_0(x)=\sum_{i=1}^{N}\gamma_i^{(0)} \nonumber\\
		\eta_i(x)=
		\begin{cases}
			\gamma_1^{(2)}-\frac{\gamma_1^{(1)}}{2}, & i=1 \\
			\gamma_i^{(2)}-\frac{\gamma_i^{(1)}}{2}+\frac{\gamma_{i-1}^{(1)}}{2}, & i=2,3,\cdots,N \\
			\frac{\gamma_N^{(1)}}{2}. & i=N+1
		\end{cases}
	\end{cases}
\end{IEEEeqnarray}

The minimum energy control is to minimize the energy consumption subject to a total trip-time constraint $ \tau $ and the optimization problem $ \mathcal{P}^{\rm ME} $ is thus formulated as:
\begin{IEEEeqnarray}{rCl}
	\begin{array}{ll}
		\underset{\substack{T_i \geq 0, x_i \geq 0}}{\rm minimize} & J^{\rm ME}(x)=\eta_0(x)+\sum_{i=1}^{N+1}\eta_i(x)x_i^2 \\
		\mbox{\rm subject to} & \sum_{i=1}^{N}T_i=\tau, ~~~~~~~~i=1,\cdots, N \\
		& T_i x_i+T_i x_{i+1}=2, ~i=1,\cdots, N \\
		& \underline{x}_i \leq x_i \leq \bar{x}_i. ~~~~~~~~i=1,\cdots, N+1
	\end{array}\nonumber
\end{IEEEeqnarray}
where $ \underline{x}_i $ and $ \bar{x}_i $ are, respectively, the lower and upper speed limits, and the traffic information can be introduced easily by adapting the speed limits to surrounding traffics. Note that $ \mathcal{P}^{\rm ME} $ involves two sets of variables, including the speed $ x $ and the travel time $ T $ to be optimized. Although the above optimization is non convex, optimizing $ x $ for fixed $ T $ can be solved easily. For fixed $ x $, solving $ T $ that satisfies the constraints amounts to solving a linear equation. Thus, a method of alternately solving $ x $ and $ T $ can be derived accordingly.

\subsection{ADMM-based optimal solution}
The alternating direction method of multipliers (ADMM) is a simple but powerful algorithm that is well suited to distributed optimization problems as described in $ \mathcal{P}^{\rm ME} $. It takes the form of a decomposition-coordination procedure, in which the solutions to small local subproblems are coordinated to find a solution to a large global problem. ADMM can be viewed as an attempt to blend the benefits of dual decomposition and augmented Lagrangian methods for constrained optimization \cite{boyd2011distributed}.

As in the method of multipliers, the augmented Lagrangian is formed by relaxing the constraints in $ \mathcal{P}^{\rm ME} $ as
\begin{IEEEeqnarray}{rCl}\label{largr}
	\tilde{L}(x,T,\mu,\lambda,\rho) &=& \eta_0(x)+\sum_{i} \eta_i(x) x_i^2+\lambda \left(\sum_{i}T_i-\tau \right)+ \sum_{i}\mu_i (T_i x_i + T_i x_{i+1} -2) \nonumber \\
	&& +\> \frac{\rho_1}{2}\left(\sum_{i}T_i - \tau \right)^2+\sum_{i}\frac{\rho_{2i}}{2}\left(T_i x_i+T_i x_{i+1}-2 \right)^2,
\end{IEEEeqnarray}
where $ \mu=[\mu_1, \mu_2,\cdots, \mu_N]^\top $ and $ \lambda $ are the Lagrange multipliers associated with constraints in $ \mathcal{P}^{\rm ME} $ and $ \rho=[\rho_1,\rho_{21},\rho_{22}, \cdots, \rho_{2N}]^\top $ are the penalty coefficients.

Minimizing the Lagrangian \eqref{largr} with respect to $ (x,T) $ is non-trivial.  The difficulty can be lessened considerably if $ x $ and $ T $ are solved separately, which gives rise to an iterative approach to solving $ \mathcal{P}^{\rm ME} $.  In particular, when the speed trajectory $ x $ is fixed, $ \eta_0 (x) $ and $  \eta_i (x) $ are determined. Eq. \eqref{largr} can thus be expressed as a quadratic function of $ T $:
\begin{eqnarray}
	\tilde{L}=T^\top A(x,\rho) T + T^\top b(x,\mu,\lambda,\rho)+c (x,T,\mu,\lambda,\rho),
\end{eqnarray}
where $ T= [T_1, T_2, \cdots, T_N]^\top$, $ A \in \mathcal{R}^{N \times N}$ is a matrix coefficient of the quadratic term, $ b \in \mathcal{R}^N $ is a vector coefficient of the first-order term, and $ c \in \mathcal{R} $ is a constant. Minimizing the Lagrangian with respect to $ T $ can be
obtained in closed form. The ADMM algorithm to solve $ \mathcal{P}^{\rm ME} $ is shown in Algorithm \ref{admm}.  

\begin{algorithm}
	\caption{The ADMM algorithm is given by the following iterations:}
	\label{admm}
	\begin{algorithmic}[1]
		\STATE Initialize $ \mu, \lambda, \rho $, the speed trajectory $ x^{(0)} $, and the trip time $ T^{(0)} $.
		\REPEAT
		\STATE Solve the primal variables:
		\begin{IEEEeqnarray}{rCl}
			x^{(n+1)} &=& \mbox{\rm argmin}\> \tilde{L} (x, T^{(n)},\mu^{(n)},\lambda^{(n)},\rho), \label{speedupdate}\\
			T^{(n+1)}&=&\mbox{\rm argmin} \>\tilde{L} (x^{(n+1)}, T,\mu^{(n)},\lambda^{(n)},\rho) \nonumber \\
			&=& -\frac{1}{2} A^{-1}(x^{n+1}, \rho) b(x^{(n+1)}, \mu^{(n)},\lambda^{(n)},\rho). \IEEEeqnarraynumspace
		\end{IEEEeqnarray}
		\STATE Update the multipliers:
		\begin{IEEEeqnarray}{rCl}
			\lambda^{(n+1)} &=& \lambda^{(n)}+\rho (\sum_{i} T_i^{(n+1)}-\tau), \\
			\mu_i^{(n+1)} &=& \mu_i^{(n)}+\rho_{2i}(T_i^{(n+1)}x_i^{(n+1)}  +T_i^{(n+1)}x_{i+1}^{(n+1)}-2).
		\end{IEEEeqnarray}
		\UNTIL{the solution difference between two iterations is small enough with $ \lVert x^{(n+1)}  - x^{(n)} \rVert_2 + \lVert T^{(n+1)} - T^{(n)} \rVert_2 \leq \varepsilon_1 $ and the constraints of the optimization problem are met with $\lVert (\sum_{i}T_i^{(n+1)}-\tau)+\sum_{i}(T_i^{(n+1)} x_i^{(n+1)}+T_i^{(n+1)} x_{i+1}^{(n+1)}-2) \rVert_2 \leq \varepsilon_2 $}, where $ \varepsilon_1 $ and $ \varepsilon_2 $ are two predefined positive real numbers close to zero.
	\end{algorithmic}
\end{algorithm}

Note that \eqref{speedupdate} that updates $ x^{(n)} $ to $ x^{(n+1)} $ can either solved by a standard solver or, perhaps more easily for a large problem, by first order (gradient) updates. 

Also Note that the sign of track force will not change suddenly as the speed trajectory $ x $ gradually changes. A positive track force within road segment $ i $ changes gradually as that in Case III (Appendix \ref{app:output}) before finally turns into a negative track force. In other words, the sign change of track force for each road segment actually represents the move of zero track force point (See Fig. \ref{segdiv} in Appendix \ref{app:output}). The continuity of track force determines the continuity of energy consumption in feasible solutions. 

Suppose there is a feasible solution $ x_0 $ and the vector parameter $ \gamma_i(x_0) $ is determined accordingly, there will be $ \lim_{x \to x_0}\gamma_i(x)=\gamma_i(x_0) $. Because $ \eta_i(x) $ consists of $ \gamma_i(x) $, there will also be $ \lim_{x \to x_0}\eta_i(x)=\eta_i(x_0)$. Since $ x_0 $ can be any feasible solution, the objective function is continuous in the feasible solutions of $ x $, which guarantees the local convergence of the proposed algorithm. In our simulation, it is observed that convergence happens approximately (or on average) in 50-100 iterations.

\subsection{MPC-based vehicle speed control}
\begin{figure*}[h]
	\centering
	\includegraphics[width=6 in]{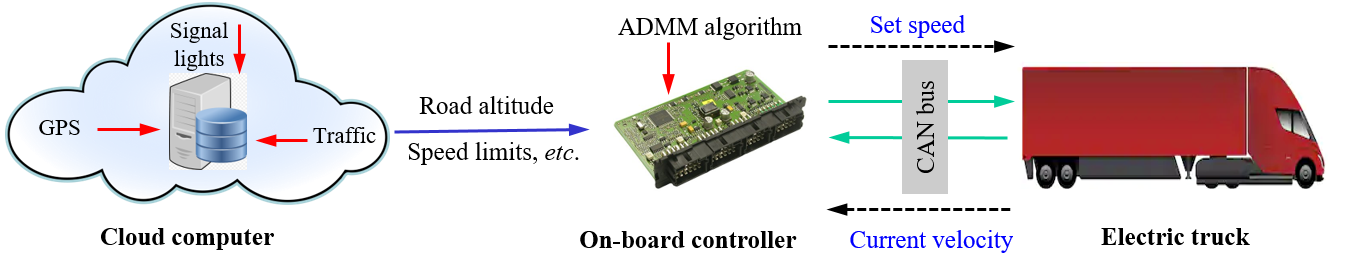}
	\caption{Information flow of the MPC framework}
	\label{mpctheo}
\end{figure*}

Model predictive control (MPC) is a rolling-window closed-loop control that incorporates real-time operating conditions. For optimal BE HDT control, MPC solves an $ N $-segment open-loop control and implements only the control of the first segment \cite{camacho2013model}.  Fig. \ref{mpctheo} shows the information flow of the MPC framework. A cloud-based platform for traffic information and situational awareness sends updated information to the on-board controller, including road altitudes, speed limits, accidents and emergencies ahead, \textit{etc}.  Based on the traffic information from the cloud platform and local sensing results (such as the distance of the car in front of the truck), the embedded ADMM algorithm calculates the optimal velocity and trip time within a small number of limited road segments, say $ N=30 $.  Only the velocity of the first segment is executed in truck operation. Note that the accurate road topographies and speeds of the preceding vehicle required by the on-board controller in this research are assumed to be available. The road grades can be estimated by fusion of GPS and vehicle real-time data, with measurements from previous runs over the same road segment \cite{sahlholm2007sensor}. The preceding vehicle trajectory can be planned or computed by the preceding vehicle and communicated to the following one \cite{dib2014optimal,xu2018cooperative,guo2018fuel,held2018optimal} or they can be computed by the following vehicle itself by exploring past and present measures of the distance and relative speed collected by on-board sensors \cite{ye2018prediction,moser2017flexible}.

\section{Case study} \label{sec: case}
\begin{figure}[h]
	\centering
	\includegraphics[width= 4 in]{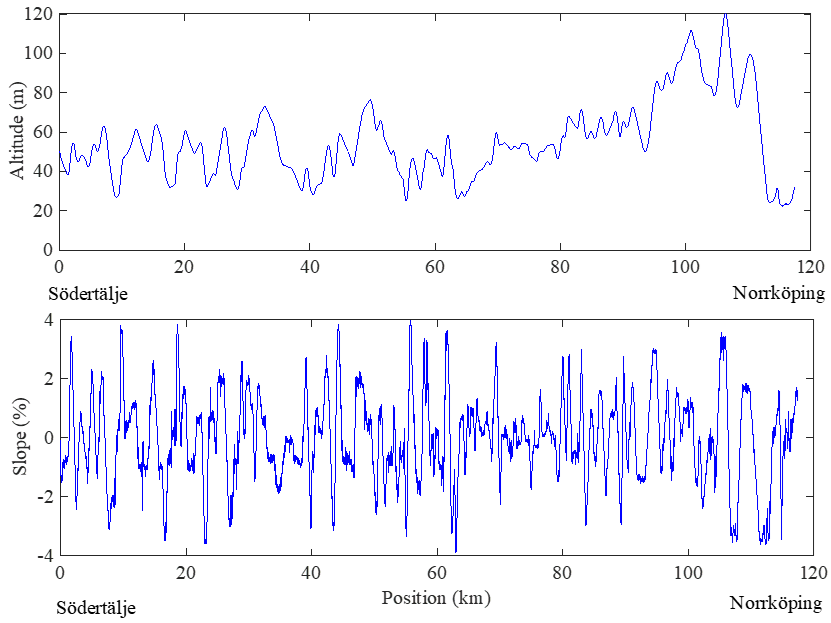}
	\caption{Estimated road topography (altitude and slope) from Södertälje to Norrköping.}
	\label{roadslope}
\end{figure}

Simulations were performed on the road data of highway E4 between the cities of Södertälje and Norrköping in Sweden \cite{hellstrom2009look}. The road slope and altitude are shown in Fig. \ref{roadslope}. The electric truck modeled was a Tesla Semi tractor and trailer. The specifications of the truck and battery pack \cite{earl2018analysis} are given in Table \ref{truckpar}. Note that the truck had four separate motors to drive the front four wheels individually, and each motor was powered by a battery pack with the nominal voltage and capacity being, respectively, 800 V and 312.5 Ah.

\begin{table}[b]
	\begin{threeparttable}
		\caption{The specifications of the Tesla Semi truck and battery pack.}
		\label{truckpar}
		\setlength\tabcolsep{0pt}
		\begin{tabular*}{\columnwidth}{@{\extracolsep{\fill}}ccc}
			\toprule
			Parameters & Description & Value \\
			\midrule
			$ m $ & Vehicle mass &  40,000  kg \\
			$ C_r $ & Rolling resistance coefficient with tyre type & 0.0055 \\
			$ A_f $ & The frontal area & 10 $ {\rm m}^2 $ \\
			$ C_D $ & Aerodynamic drag coefficient & 0.36 \\
			$ E_T $ & Nominal energy in total & 250 kWh *4 \\
			$ U $ & Nominal voltage (one battery pack) & 800 V \\
			$ C $ & Nominal capacity (one battery pack) & 312.5 Ah \\
			$ \beta^+ $ & Vehicle discharge efficiency (battery to wheel) & 0.85 \\
			$ \frac{1}{\beta^-} $ & Vehicle charge efficiency (wheel to battery) & 0.80 \\
			\bottomrule
		\end{tabular*}
	\end{threeparttable}
\end{table}

Algorithm parameters were initialized as follows. The length of each road segment $ l $ was set to 50 m, and the number of segments $ N $ was set to 30. The setting of 50 m length for each segment, which value can be adjusted according to practical conditions, has been proved to be feasible in practice \cite{hellstrom2009look, turri2018fuel}. Therefore, the prediction horizon was 1500 m. The speed lower bound was set to 0 km/h. In the case without traffics involved, the speed upper bound was set to the EU legal maximum of 90 km/h and the speed lower bound was set to 75 km/h \cite{hellstrom2009look}, while in the case under traffics, the speed was limited to surrounding traffics.
Simulations were conducted in the environment of Matlab 2020a based on a Macbook Pro with a processor of 8-Core Intel Core i9 @2.4 GHz and a memory of 32 GB.
\subsection{Optimal speed control without traffics involved}
 In this case, the trip time $ \tau $ was set equal to the trip time of that using a uniform speed of 85 km/h to travel through the same distance.
\subsubsection{Overall performance}
\begin{figure}[t]
	\centering
	\includegraphics[width=3.5 in]{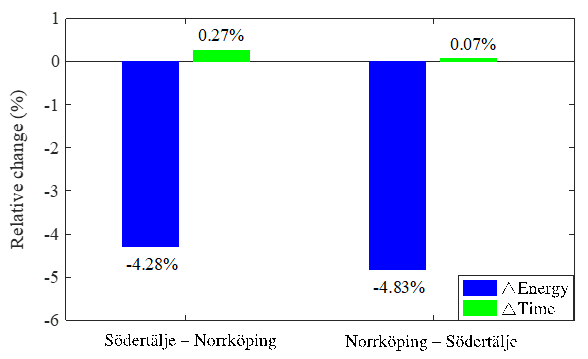}
	\caption{Simulation results on the road data from Södertälje to Norrköping and the return.}
	\label{overallperf}
\end{figure}

Two simulations were conducted based on the road data between Södertälje and Norrköping to compare the performance of the ADMM controller and the uniform speed cruise control (CC) controller.  The CC speed was set at 85 km/h, and the relative changes of energy consumption and trip time  between the two controllers are shown in Fig. \ref{overallperf}. A negative value indicates that the ADMM controller has a lower value than the CC does. The results show that, compared to the CC the ADMM controller saved 4.28\% energy from Södertälje to Norrköping and 4.83\% energy from the return, while the trip time between these two controllers were similar to each other in both directions.
\subsubsection{Performance comparison}
\begin{figure*}[h]
	\centering
	\includegraphics[width=5 in]{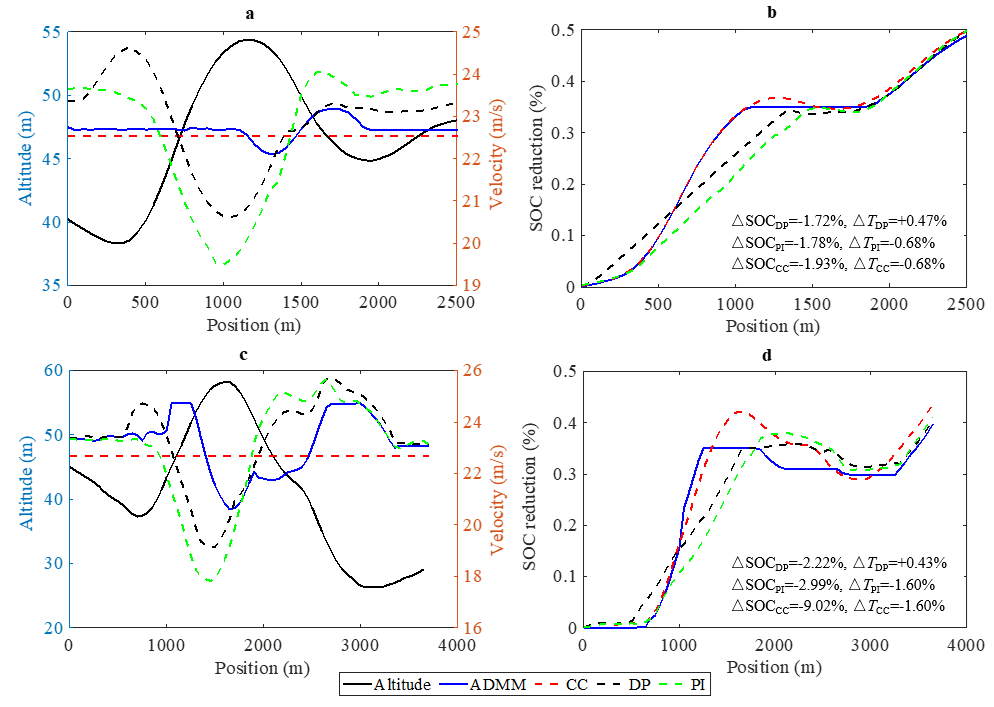}
	\caption{The optimization results of truck speed.  (a) and (c), which respectively relates to Figs. 9 and 7 in \cite{hellstrom2009look}, present the speed optimization results and the road altitude data. The corresponding energy consumption, which is indicated by SOC reduction, is presented in (b) and (d).}
	\label{compr1}
\end{figure*}

Since there is no published work on energy-efficient eco-driving of BE HDT, the comparisons were made with energy-efficient driving algorithms for traditional trucks \cite{hellstrom2009look} where dynamic programming (DP) and proportional integral control (PIC) were proposed. And the optimal speed trajectories from \cite{hellstrom2009look} are directly introduced for performance comparisons. According to \cite{hellstrom2009look}, the PIC is a standard controller available from Scania. All parameters that could affect the vehicle energy consumption were set to the same as those in \cite{hellstrom2009look}. Fig. \ref{compr1} presents the comparison results between ADMM, DP, PIC and CC based on the road slopes of Figs. 7 and 9 in \cite{hellstrom2009look}. The speed value of the CC was set to have the same trip time as that of PI. The relative changes in energy consumption and trip time ($ \Delta{\rm {SOC}} $, $ \Delta T $) of ADMM to other methods are also presented in Fig. \ref{compr1} for each road scenario.

Since the charging/discharging efficiency of the vehicle is less than 1, the regenerative braking is only able to regenerate a part of the consumed energy. Besides, moving the truck forward also consumes energy to overcome different kinds of resistances, and this energy consumption further reduces the efficiency of the regenerative braking. Therefore, energy consumption can be reduced by avoiding any undesirable braking. 

Fig. \ref{compr1}(a) shows that the HDT kept constant speed based on ADMM except during downhill stretches, where the truck decelerated first, then accelerated, finally decelerated again to a constant speed. The ADMM-based energy consumption on the downhill was close to zero as shown in  Fig. \ref{compr1}(b), which indicates the truck moved forward in the most energy conserving fashion  with  little undesirable braking,  whereas more braking events were observed in  Fig. \ref{compr1}(b) for all other methods including the DP, PIC, and CC.  Note that traditional trucks needed to downshift (decelerate) to increase the driving force when going uphill (see DP and PIC-based speed trajectory in  Fig. \ref{compr1}(a)), which caused the undesirable acceleration and braking on the downhill (see DP and PIC-based energy consumption in  Fig. \ref{compr1}(b)) to meet the trip time requirement.  Because the BE HDT powered by the motor was able to go uphill without decelerating and with a high speed, it left wider improvement space for improving energy consumption when going downhill than the traditional truck did. This phenomenon, which has never been revealed in the literature, indicates the BE HDT adapts to the road topography in a more energy efficient way than the ICE HDT does. In each case, the CC performed the worst since it used the most undesirable braking and thus consumed the most energy to keep a constant speed when going downhill. The energy consumption results in Fig. \ref{compr1}(b) show that, with a similar trip time, the ADMM-based BE HDT, respectively, consumed 1.72\% and 1.78\% less energy than the DP- and PIC-based ICE HDT, and 1.93\% less energy than the CC-based BE HDT.

Fig. \ref{compr1}(c) shows the optimal speed trajectories on a road with a long downhill segment where braking was inevitable for the BE HDT within the speed upper limit.  In this case, the ADMM used less braking and thus consumed less energy than other methods did (Fig. \ref{compr1}(d)). Obviously, the driving characteristics of BE HDT was quite different from those of the ICE-powered HDT. The energy consumption results show that the ADMM-based BE HDT, respectively, consumed 2.22\% and 2.99\% less energy than the DP- and PIC-based ICE HDT, and 9.02\% less energy than the CC-based BE HDT while keeping a similar trip time. Note that the reason that ADMM saves more energy than DP in the two cases is mainly due to the different vehicle types (i.g. BE HDT vs. ICE HDT). The energy consumption values are expected to be similar if the two controllers are both developed for the BE HDT.

\subsubsection{Computation time}
An important issue concerning optimization algorithms for speed trajectory planning is the real-time capability. Considering the implementation in a truck, the algorithm must be able to adapt to upcoming uncertainties such as the traffic flow and signal variations. Within the speed limit, the shortest trip time for the truck to go through one road segment is 2 s. Thus to realize real-time optimal planning of the speed trajectory, a maximum turnaround time of 2 s is desired.

Simulations were, respectively, conducted for 10 times based on road data in Figs. \ref{compr1}(a) and (c). The averaged computation time for planning the speed trajectory over the road in Fig. \ref{compr1}(a) is 53.2 s with a maximum value of 53.9s. The computation time for each road segment is  thus 1.06 s on average considering there are 50 road segments for this scenario. The road in Fig. \ref{compr1}(c) contains 70 road segments, and the averaged computation time required for speed planning is 72.6 s with a maximum value of 72.8 s. Thus the average computation time for each road segment is 1.04 s. In practice, structured optimization algorithms can be specifically designed for solving the optimal control problem more efficiently, leading to on-board controllers with promising real-time capability.

\subsection{Optimal speed control under traffics}
As indicated before, the surrounding traffics can be introduced easily to the optimization $ \mathcal{P}^{\rm ME} $. Appendix \ref{app:traff} shows the detailed modeling processes where the vehicle drives in a safe and energy-efficient way, and specifically, the upper-speed limit $ \bar{x}_i $ is adapted to the speed of the preceding vehicle.
\subsubsection{Traffic stochastic modeling}
Traffics around the truck need to be simulated for traffic-based optimal speed control. Herein, the exponential distribution was used to generate the stochastic traffic. The exponential distribution is a standard distribution that can model the inter-arrivals between vehicles \cite{balakrishnan2019exponential}. Let event ‘1’ and ‘0’ represent, respectively, that there is a preceding vehicle or not. The traffic is a Markov process if the exponential distribution is used to represent the elapsed time between event ‘1’ and ‘0’ \cite{ethier2009markov}. A continuous random variable $ X $ is said to have an exponential distribution if it has probability density function
\begin{IEEEeqnarray}{rCl}
	f_X(x|\lambda)=
	\begin{cases}
		\lambda e^{-\lambda x}, & x>0 \\
		0, & x \leq 0 \nonumber
	\end{cases}
\end{IEEEeqnarray}
where $ \lambda>0 $ is called the rate of the distribution. Herein, $ X $ represents the traveling distance instead of the duration time as the distance and the duration time can always transform between each other in this case. $ X_1 $ is used to represent the traveling distances with preceding vehicles and $ X_2 $ to represent the traveling distances without preceding vehicles, and the mean values of these two distributions are thus $ \mu_1=\frac{1}{\lambda_1} $  and $ \mu_2=\frac{1}{\lambda_2} $ . The traffic situation is decided by the values of $ \mu_1 $ and $ \mu_2 $. A large $ \mu_1 $ and a small $ \mu_2 $ indicate heavy traffic and vice versa. $ \mu_1 $ and $ \mu_2 $ can be calibrated based on available real traffic flow data.

\subsubsection{Model initialization}
Some model parameters need to be initialized for defining the traffic characteristics. In this study, the minimum headway $ h_\tau $ from the preceding vehicle was set to 1.2 s for the autonomous following trucks to ensure traffic safety \cite{nodine2017naturalistic}. The initial distance $ d_i $ from the preceding vehicle was assumed within [2, 4] s headway with uniform distribution, and the speed of the preceding vehicle $ v_p $ was assumed within [70, 80] km/h with uniform distribution. Heavy traffic is characterized by ($ \mu_1=3,\mu_2=2 $); Light traffic is characterized by ($ \mu_1=2,\mu_2=3 $), and normal traffic is characterized by ($ \mu_1=3,\mu_2=3 $). Note that in this case, the trip time $ \tau $ is highly dependent on the traffic situations and thus cannot be determined in advance.

\subsubsection{Speed control results with traffics involved}
In this part, we first investigated the energy consumption results for trucks following a preceding vehicle on a flat road (road slope = 0). Then we further investigated the truck energy consumption following a preceding vehicle on a road with varying slopes. Finally, the truck energy consumption results from Södertälje to Norrköping and the return were evaluated based on heavy and light traffics, respectively. In the scenarios with surrounding traffics involved, since no DP-or PIC-based speed trajectory can be introduced for comparisons, only the CC-based speed trajectory was used as a benchmark.

\begin{figure}[h]
	\centering
	\includegraphics[width=4.5 in]{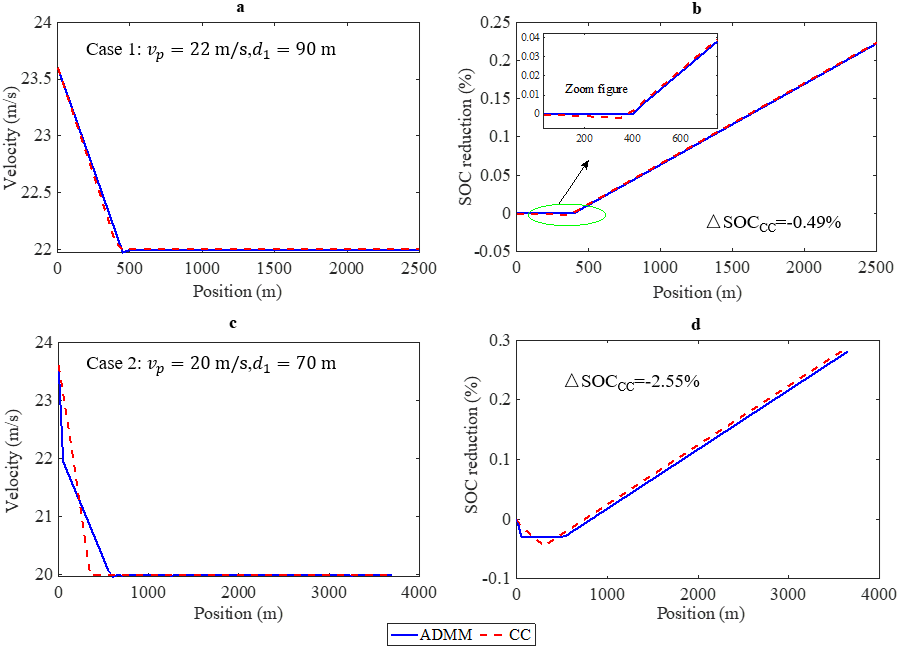}
	\caption{Simulation results based on flat roads: (a) Speed trajectories based on ADMM and CC with a preceding vehicle velocity of 22 m/s and initial spacing of 90 m, and SOC reductions as the position are presented in (b); (c) Speed trajectories based on ADMM and CC with a preceding vehicle velocity of 20 m/s and initial spacing of 70 m, and SOC reductions as the position are presented in (d).}
	\label{trafficflat}
\end{figure}

Fig. \ref{trafficflat} presents the simulation results on flat roads with different velocities of the preceding vehicle and initial distances.  Note that the CC speeds were obtained based on two rules: first, the CC travel time was the same as that based on ADMM, and second, the headway for CC was also set to 1.2 s. It was observed that the ADMM controller saved more energy than the CC did in both cases because the ADMM controller braked less on flat roads. In case 1, the ADMM used resistances (air and frictional) to decelerate while the CC used additional braking which thus consumed more energy. In case 2, the ADMM controller braked for a little while at first then decelerated perfectly depending on resistances, whereas the CC controller braked until reaching the desired speed. Fig. \ref{trafficflat} indicates that our method still worked on saving energy for trucks following a preceding vehicle, and more energy was expected to be further saved with the road slopes involved.

Figure 7 presents the simulation results with road slopes from Fig. \ref{compr1}. It was observed that the ADMM still braked less and was more energy efficient. Figure 7 (b) and (d) show the energy consumption based on ADMM was, respectively, reduced by 6.07\% and 19.46\% compared to that based on CC. These results indicate that our method still performed excellently on energy consumption improvement for trucks following a preceding vehicle on roads with varying slopes.
\begin{figure}[h]
	\centering
	\includegraphics[width=4.5 in]{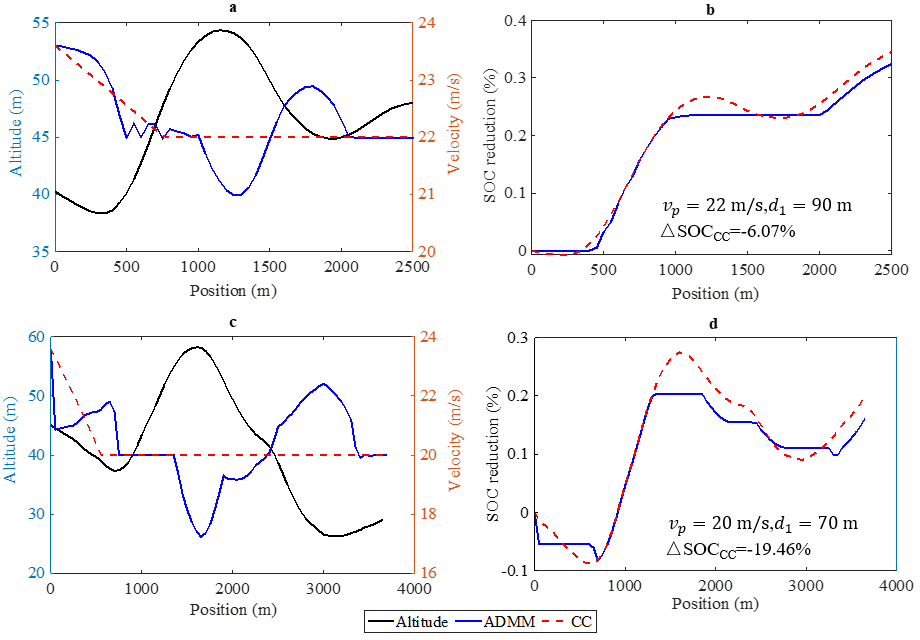}
	\caption{Simulation results with both road slopes and traffics involved: (a) Speed trajectories and road altitude corresponding to Fig. \ref{compr1} (a); (b) SOC reductions; (c) Speed trajectories and road altitude corresponding to Fig. \ref{compr1} (c); (d) SOC reductions.}
	\label{trafficslo}
\end{figure}

\begin{figure*}[h]
	\centering
	\includegraphics[width=2.5 in]{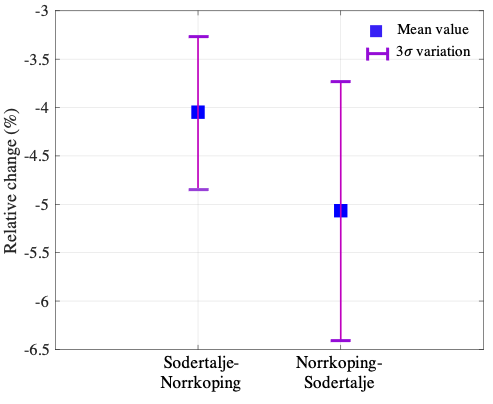}
	\caption{Simulation results of energy consumption under stochastic traffics from Södertälj to Norrköping and the return.}
	\label{energysav}
\end{figure*}

Fig. \ref{energysav} shows the statistical results under stochastic traffics from Södertälj to Norrköping and the return. For each direction, 10 traffic scenarios covering heavy ($ \mu_1=3,\mu_2=2 $), light ($ \mu_1=2,\mu_2=3$) and normal ($ \mu_1=3,\mu_2=3 $) traffics were generated randomly and implemented to verify the performance of ADMM. For each traffic scenario, the preceding vehicle turns in and out randomly, and the total driving ranges with a preceding vehicle (labeled 'Traffic') and without any preceding vehicles (labeled 'No traffic') are listed in Table \ref{tab:traffic}. From Södertälj to Norrköping, Scenarios 1 to 3 indicate heavy traffic; Scenarios 4 to 7 indicate light traffic while Scenarios 8 to 10 indicate normal traffic.  For the return direction, Scenarios 1 to 4 indicate heavy traffic; Scenarios 5 to 7 indicate light traffic while Scenarios 8 to 10 indicate normal traffic. 

Under each traffic scenario, the relative change of energy consumption between ADMM and CC was calculated, and the mean values and $ 3\sigma $ variations were further obtained by summarizing the simulation results under different traffic scenarios. It was observed that from Södertälj to Norrköping, the energy consumption of ADMM was reduced by 4.05\% with a $ 3\sigma $ interval being [3.27\%, 4.83\%] compared to that based on CC, while from Norrköping to Södertälj, the saved energy was up to 5.07\% with a $ 3\sigma $ interval being [3.73\%, 6.40\%]. These results of energy saving indicate that our method adjusts well to different traffic situations for minimal energy consumption purposes.
\begin{table}
	\caption{Details of 10 traffic scenarios.}
	\label{tab:traffic}
	\begin{tabular}{m{5em}lllllllllll}
		\toprule
		Direction& {\color{red}Label}&{\color{red}S1(km)}&{\color{red}S2(km)}&{\color{red}S3(km)}&{\color{red}S4(km)}&{\color{red}S5(km)}&{\color{red}S6(km)}&{\color{red}S7(km)}&{\color{red}S8(km)}&{\color{red}S9(km)}&{\color{red}S10(km)}\\
		\midrule
		\multirow{2}{5em}{Södertälj to Norrköping} & Traffic &77.7&69.75&70.65&44.7&48.3&40.15&42.7&49.1&56.45&51.35\\
		&No traffic &35.6&41.4&42.2&68.85&63.65&73.3&71.65&63.95&55.0&63.75\\
		\multirow{2}{5em}{Norrköping to Södertälj}& Traffic &72.7&61.5&69.7&78.45&35.5&45.2&40.15&58.15&55.55&56.65\\
		&No traffic &40.9&49.15&44.45&35.8&80.5&67.4&72.85&59.2&61.35&60.65\\
		\bottomrule
	\end{tabular}
\end{table}

\section{Evaluation of Battery Aging under Eco-driving}
Energy minimization is only one of issues of BE HDT eco-driving control. Also relevant is the impact of driving algorithm on battery health. Limited to the current battery technology, the battery life, which is generally shorter than the EV life \cite{yang2018predictive}, is even shorter for BE HDT owing to the long-haul driving requirement. Therefore, battery aging in BE HDT  needs not only to be on-line monitored but also to be further improved.

Indeed, many studies have been conducted on battery SOH estimation in recent years, e.g., Chen et al. \cite{chen2017lithium} developed a promising method by using dual H infinity filters to estimate battery SOH in real-time and the high estimation accuracy was verified by the hardware-in-loop experiment, while there is limited literature on managing EV real-time operations for extending battery life. It is demonstrated here the positive effect of the proposed approach on battery aging.

\subsection{EV battery life model}
Battery aging includes cycling aging and calendar aging \cite{zhang2019state, zhang2019aging, lam2012practical}. For a long-haul BE HDT, one suitable operation mode would be shipping during the day and charging at night. This operation mode indicates that the truck battery pack will be cycled most of the day. In this case, it is only needed to evaluate the cycling aging of batteries for the BE HDT. 

The battery cycling aging model used in this paper is developed in \cite{lam2012practical}. This model is selected for two reasons: first, this model analyses the effect of the regenerative braking on battery aging and involves this effect in the aging model. Second, the model is verified by a large amount of experimental data simulating EV operations. 
The total capacity fade at a constant temperature is expressed as
\begin{IEEEeqnarray}{rCl}
	\xi=\xi'({\rm SOC}_{\rm avg}, {\rm SOC}_{\rm dev}) \times Q,
\end{IEEEeqnarray}
where $ \xi $ is the total capacity fade, $ Q $ is the ampere-hour (Ah) charge processed during charging/discharging, and $ \xi' $ is the capacity fading rate and its unit is Ah faded per Ah processed by batteries. The capacity fading rate is expressed as
\begin{IEEEeqnarray}{rCl}
	\xi'=k_1 {\rm SOC}_{\rm dev} e^{k_2 {\rm SOC}_{\rm avg}}+k_3 e^{k_4 {\rm SOC}_{\rm dev}},
\end{IEEEeqnarray}
where $ k_1 $ to $ k_4 $ are four model fitting parameters. $ {\rm SOC}_{\rm avg}=\frac{1}{Q_{t1-t0}}\int_{Q_{t0}}^{Q_{t1}}{\rm SOC}(Q) \, dQ $ indicates the average SOC, with $ Q_{t0} $ being the amount of charge processed at the moment $ t0 $, $ Q_{t1} $ being the amount of charge processed at the moment $ t1 $, and $ Q_{t1-t0}=Q_{t1}-Q_{t0} $. $ {\rm SOC}_{\rm dev}=\sqrt{\frac{1}{Q_{t1}-Q_{t0}}\int_{Q_{t0}}^{Q_{t1}}({\rm SOC}(Q)-{\rm SOC}_{\rm avg})^2 \, dQ} $ indicates the normalized standard deviation from $ SOC_{\rm avg} $. It is assumed the thermal management system of vehicle is able to keep the battery pack at a constant temperature of \SI{25}{\celsius} and model parameters $ k_1 $ to $ k_4 $ are thus initialized at this specific temperature \cite{lam2012practical}.

\subsection{Optimal speed control impact on battery aging}
\subsubsection{Parameter pre-setting of simulation}
The battery aging model indicates that, in general, the battery life is longer within the SOC cycling range of a lower SOC average value. Therefore, the ending SOC instead of the initial SOC is adjusted to the same value for each controller for fair comparisons. Generally, a long-haul HDT should run more than 500 km each working day. The road data from Södertälj to Norrköping and the return (about 240 km long) was thus repeated to generate the road data profile with the wanted travel distance. It was assumed the truck shipped during the day and got charging at night. A slow charging was preferred for extending the battery life and thus the charging rate was set at 0.1 C in simulation. Note that the truck was expected to be used each working day and thus it would operate 260 days each year. Currently, the EV battery degradation limit is agreed upon 30\% limit \cite{yang2018predictive}.

\subsubsection{Battery aging without traffics involved}
 Herein three cases were simulated where for each case the truck drove a pre-determined distance, or the battery pack reached a pre-determined ending SOC each day.

\textit{Case 1:} The truck traveled 800 km long each day, which is the Tesla claimed truck driving range with batteries fully charged. For fresh new battery packs, the battery (depths of discharge) DODs for the ADMM and CC benchmark were, respectively, 90.86\% and 95.12\% at the end of trip and the Ah throughputs of one battery pack were, respectively, 328.8 Ah and 400.3 Ah. It was observed that, after speed control using ADMM, the charge delivered by battery was reduced by 71.5 Ah, which accounted for 22.9\% of the battery nominal capacity (312.5 Ah). As the capacity degradation is proportional to the Ah throughput (see battery aging model), the ADMM controller was expected to reduce battery aging by more than 20\% compared to the CC policy. This significant improvement showed that, when compared with the CC policy, the proposed eco-control algorithm not only minimized energy consumption but also extended battery life. 

The improvement of battery health came from the fact that the proposed minimum energy control avoided undesirable braking. Note that the battery delivered about twice the regenerative charge by each undesirable braking comparing with the case when undesirable brakings were avoided.

In this case, the SOC cycling ranges for ADMM and CC benchmark were, respectively, [95.74\%, 4.88\%] and [100\%, 4.88\%]. The battery aging evaluation results are presented in Table \ref{case1}. It was surprising to find that the one year-capacity fading $ \xi $ of battery  based on ADMM was reduced by 35.2\% compared to CC. The ADMM controller led to lower battery DOD and thus smaller ${\rm SOC}_{\rm avg} $ and $ {\rm SOC}_{\rm dev} $ than the CC controller did. This low battery DOD caused a reduction of 15.5\% on the capacity fading rate $ \delta \xi $ using ADMM compared to using CC, which explains why ADMM  reduced the capacity degradation of battery much higher than the afore-mentioned value of 20\%.
\begin{table*}[t]
	\centering
	\caption{Battery aging and life estimation results for case 1.}
	\label{case1}
	\begin{threeparttable}
		\begin{tabular}{c*{4}{c}}
			\toprule
			& $ {\rm SOC}_{\rm avg} $ & $ {\rm SOC}_{\rm dev} $ & $\delta \xi (\times 10^{-4} {\rm Ah})$ & $ \xi $ (\%, one year) \\
			\midrule
			CC & 0.53 & 0.47 & 1.94 & 11.30 \\
			ADMM & 0.50 & 0.45 & 1.64 & 8.36 \\
			\bottomrule
		\end{tabular}
	\end{threeparttable}
\end{table*}

\begin{figure}[h]
	\centering
	\includegraphics[width=3 in]{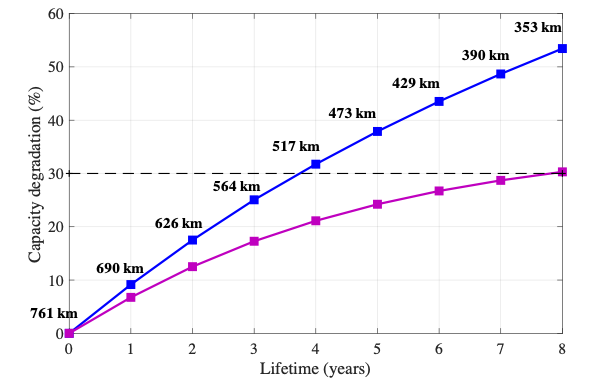}
	\caption{\color{red}Battery aging and life estimation results for Case 2 of ADMM (dark red) and CC (blue).}
	\label{fig:case2}
\end{figure}

\begin{figure}[h]
	\centering
	\includegraphics[width=2.8 in]{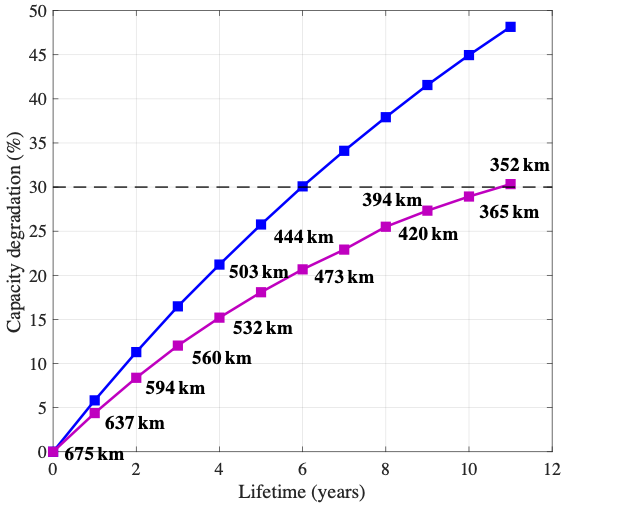}
	\caption{\color{red}Battery aging and life estimation results for Case 3 of ADMM (dark red) and CC (blue).}
	\label{fig:case3}
\end{figure}

\textit{Cases 2 and 3:} Generally, an EV needs to be recharged when the battery SOC is lower than 10\%-20\%. Therefore, in Cases 2 and 3, the battery ending SOC for CC was, respectively, setting to 10\% and 20\%, which corresponds to a CC-based truck travel distance of 761 km and 675 km each day for fresh new batteries. For fair comparisons, the driving range based on ADMM was set to the same value as that based on CC. Fig. \ref{fig:case2} and Fig. \ref{fig:case3} show the battery aging results for Case 2 and Case 3, respectively, where the CC-based driving range within the corresponding DOD is also presented. It was observed that the driving range decreased as the battery capacity degraded. In Case 2 (Fig. \ref{fig:case2}), the CC- and ADMM-based battery lives were 3.74 years and 7.81 years long, respectively, indicating a life extension of 108.8\% for the ADMM. At the battery end of life (EOL) for CC, the truck driving range decreased to about 517 km, equal to a range shrinkage of 32\% compared to the initial driving range.

In Case 3 (Fig. \ref{fig:case3}), because the battery DOD was 10\% shorter than that in Case 2, the CC- and ADMM-based battery lives were both longer and were about 6 years and 10.5 years long, respectively, indicating a life extension of 75\% for ADMM. In this case, the reduction of driving range was equal to 202 km, amounts to a range shrinkage of up to 43\%, which indicated the necessity to reduce battery aging from EV driving perspectives.

\subsubsection{Battery aging under traffics}
\begin{figure}[h]
	\centering
	\includegraphics[width=3 in]{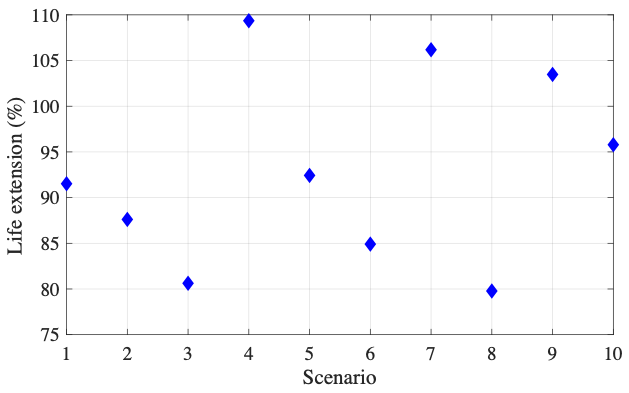}
	\caption{ADMM-based life extension results under traffics.}
	\label{fig:traffic}
\end{figure}

In this case, the optimal speed profiles under traffics were used, and as described before the generated stochastic traffic scenarios from Södertälje to Norrköping and the return covered heavy, light and normal traffics. For each scenario, the road altitudes and traffics from Södertälj to Norrköping and the return were connected and repeated to generate the desired road profiles for simulation, leading to totally 10 simulated situations under different traffics. Fig. \ref{fig:traffic} shows the ADMM-based life extension results compared to CC for these 10 traffic scenarios, where the CC-based ending SOC each day was set to 15\% over different aging stages. The ADMM-based driving range each day was also set to the same value as that based on CC for fair comparisons. It was observed that the ADMM-based life extension ranged from 80\% to 110\%, with a mean value of up to 93.2\%.

\section{Conclusion}
This paper has developed a methodology of controlling truck speed with minimal energy consumption and extended battery life. The novel state-space equations are constructed to describe the system dynamics of the truck. The dependencies of truck operation speed and energy consumption are captured by a state-space model with truck speed as the state and battery state of charge (SOC) as the output. An energy minimization problem is defined, where a novel optimization technique based on the principle of alternating direction method of multipliers (ADMM) is introduced for optimal solutions, coupled with a model predictive control (MPC) strategy to deal with the uncertainty of the upcoming road topography and traffic for planning the truck speed in real-time. The performance of the developed method is verified based on real highway altitudes between the cities of Södertälj and Norrköping in Sweden.

The simulation results show that the developed method is able to exploit topographical conditions for improved energy management, both in terms of minimizing total battery discharge and prolonging battery lifetime. It shows that the ADMM control consumes less energy under different scenarios than the dynamic programming (DP) control, proportional integral control (PIC) and uniform speed cruise control (CC). Generally, ADMM consumes 4\%-5\% less energy than CC does. It is surprising to find that ADMM generally extends battery life by more than 1 time than CC does.These results suggest the necessity to improve battery energy consumption and aging by optimizing truck speed trajectories. The battery energy and aging improvement values still hold when the traffic is introduced, which indicates that our method is also able to be used to electric buses and passenger electric vehicles in urban driving.


%

\appendices
\section{Electric motor modeling} \label{app:EM}
\begin{figure}[h!]
		\centering
		\includegraphics[width=3 in]{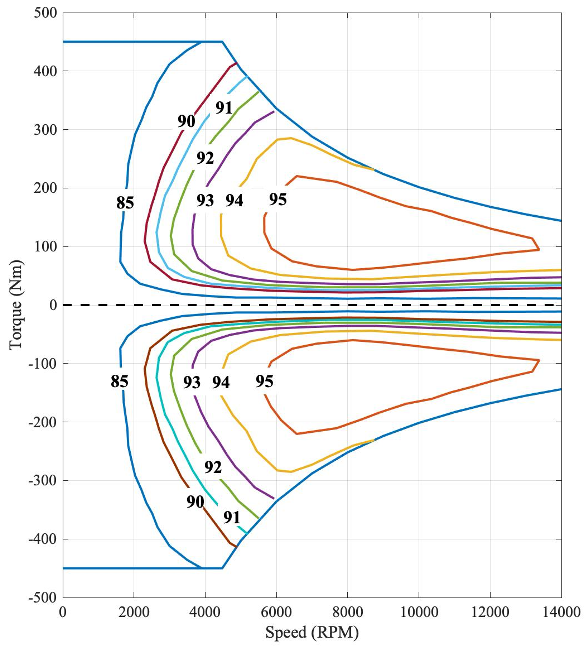}
		\caption{\color{red}EM efficiency map of Tesla Semi.}
		\label{fig:EMmap}
	\end{figure}

	The electric motor (EM) is assumed to have similar efficiencies under traction and regenerative braking modes. The EM power $ P^{\rm EM} $ is expressed as:
	\begin{eqnarray}
		P^{\rm EM}=
		\begin{cases}
			\frac{\pi n_m T^{\rm EM}}{30 \eta(n_m, T^{\rm EM})},~~~~~{\rm driving} \\
			\frac{\pi n_m T^{\rm EM} \eta(n_m, T^{\rm EM})}{30},~~~{\rm braking}
		\end{cases} \nonumber
	\end{eqnarray}
	where $ T^{\rm EM} $ indicates the EM torque, $ n_m $ indicates the rotational speed of the EM, and $ \eta(n_m, T^{\rm EM}) $ indicates the motor efficiency depending on $ n_m $ and $ T^{\rm EM} $. The EM efficiency map of the Tesla Semi is presented in Fig. \ref{fig:EMmap} \cite{motor2020performance}.
	
	\textit{Remark:} It was claimed that the Tesla Semi shares a number of parts with its Model 3, including the same motor
	\cite{alex2017meet} (Fig. \ref{fig:EMmap}).
	
	The vehicle speed $ v $ and the EM rotational speed $ n_m $ is connected by
	\begin{eqnarray}
		n_m=i_g \frac{30 v}{\pi r_{\rm whl}}, \nonumber
	\end{eqnarray}
	where $ i_g $ indicates the gear ratio, and $ r_{\rm whl} $ indicates the radius of the vehicle wheel. Note that the truck speed on highway is limited within [75, 90] km/h \cite{hellstrom2009look}. In this case, the rotational speed range of EM is within [7000, 9000] rpm at a gear ratio of 19:1 \cite{keith2018exploring}. Within this rotational speed range, the EM operates at a highly efficient mode with efficiencies being around 94\%. Therefore, the EM efficiency can also be considered a constant when calculating the powertrain efficiency $ \beta_i $.

\section{Output function modeling} \label{app:output}
Four cases are considered herein.

\textbf{Case I:} $ F_{i}^{\rm track}(t) > 0 $ within segment $ i $, \emph{i.e.}, the battery always discharges. Note that since the term $ \frac{1}{UC} $ in \eqref{socch} is a constant, which will be removed temporarily for concise expression in the following derivations but will be added back in the simulation results.
\begin{IEEEeqnarray}{rCl}
	G_{i}(x_i,a_i) &=& \int_{0}^{T_{i}}p_{i}(t) \,dt = \int_{0}^{T_i}\frac{1}{\beta^+}v_{i}(t)F_{i}^{\rm track}(t)\,dt \nonumber\\
	&=& \frac{ma_{i}+\beta_{i}^{(0)}}{\beta^+}\int_{0}^{T_i}(x_{i}+a_{i}t)\,dt+\frac{\beta^{\rm air}}{\beta^{+}}\int_{0}^{T_{i}}(x_{i}+a_{i}t)^{3}\,dt \nonumber\\
	&=& \frac{ma_{i}+\beta_{i}^{(0)}}{2a_{i}\beta^{+}} \left[\left(x_{i}+a_{i}T_{i}\right)^{2}-x_{i}^{2}\right] + \frac{\beta^{\rm air}}{4a_{i}\beta^{+}}\left[\left(x_{i}+a_{i}T_{i}\right)^{4}-x_{i}^{4}\right] \nonumber\\
	&=& \frac{1}{\beta^+}\left(ma_{i}+\beta_{i}^{(0)}\right)+\frac{\beta^{\rm air}}{\beta_{+}}\left(x_{i}^{2}+a_{i}\right) \nonumber \\
	&=& \gamma_{i}^{(0)}+\gamma_{i}^{(1)}a_{i}+\gamma_{i}^{(2)}x_{i}^{2},
\end{IEEEeqnarray}
where $ \gamma_{i}^{(0)}=\frac{\beta_{i}^{(0)}}{\beta^{+}}, \gamma_{i}^{(1)}=\frac{m+\beta^{\rm air}}{\beta^{+}}, \gamma_{i}^{(2)}=\frac{\beta^{\rm air}}{\beta^{+}} $.

\textbf{Case II:} $ F_{i}^{\rm track}(t) < 0 $ within segment $ i $, \emph{i.e.}, the battery always charges. The derivation for Case I applies. Substituting $ \beta^{+} $ by $ \beta^{-} $, there is
\begin{IEEEeqnarray}{rCl}
	G_{i}(x_i,a_i) = \gamma_{i}^{(0)}+\gamma_{i}^{(1)}a_{i}+\gamma_{i}^{(2)}x_{i}^{2}, \nonumber
\end{IEEEeqnarray}
where $ \gamma_{i}^{(0)}=\frac{\beta_{i}^{(0)}}{\beta^{-}}, \gamma_{i}^{(1)}=\frac{m+\beta^{\rm air}}{\beta^{-}}, \gamma_{i}^{(2)}=\frac{\beta^{\rm air}}{\beta^{-}} $.

\textbf{Case III:} The sign of $ F_{i}^{\rm track} $ changes from positive to negative, \emph{i.e.}, there exists $ t_{i}^{*} $ such that
\begin{IEEEeqnarray}{rCl}
	\begin{cases}
		F_{i}^{\rm track}(t) \geq 0, \; t_{i} \leq t \leq t_{i}^{*} \nonumber\\
		F_{i}^{\rm track}(t) \leq 0. \; t_{i}^{*} \leq t \leq t_{i+1} \nonumber
	\end{cases}
\end{IEEEeqnarray}

\begin{figure}[h]
	\centering
	\includegraphics[width=2.5 in]{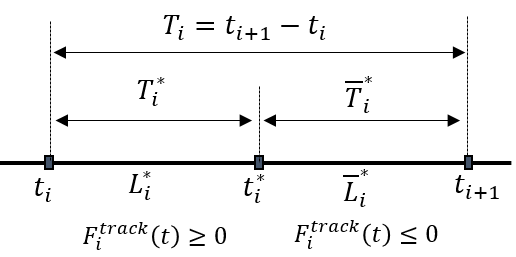}
	\caption{Schematic diagram: the track force is positive before $ t_{i}^{*} $ and then changes to negative after this moment.}
	\label{segdiv}
\end{figure}

Note that $ t_{i}^{*} $ is a function of $ (x_i,a_i) $. The schematic diagram of this case is shown in Fig. \ref{segdiv}

The output function is the sum of two sub-segments
\begin{IEEEeqnarray}{rCl}
	G_{i}(x_i,a_i)=G_{i}^{+}(x_i,a_i)+G_{i}^{-}(x_i,a_i), \nonumber
\end{IEEEeqnarray}
where
\begin{IEEEeqnarray}{rCl}\label{output3}
	G_{i}^{+}(x_i,a_i) &=& \int_{t_i}^{t_{i}^{*}}p_{i}(t) \,dt =
	\frac{ma_{i}+\beta_{i}^{(0)}}{\beta^+}\int_{0}^{T_{i}^{*}}(x_{i}+a_{i}t)\,dt + \frac{\beta^{\rm air}}{\beta^{+}}\int_{0}^{T_{i}^{*}}(x_{i}+a_{i}t)^{3}\,dt \nonumber\\
	&=& \frac{ma_{i}+\beta_{i}^{(0)}}{2a_{i}\beta^{+}} \left[\left(x_{i}+a_{i}T_{i}^{*}\right)^{2}-x_{i}^{2}\right] + \frac{\beta^{\rm air}}{4a_{i}\beta^{+}}\left[\left(x_{i}+a_{i}T_{i}^{*}\right)^{4}-x_{i}^{4}\right],
\end{IEEEeqnarray}
where $ T_{i}^{*}=t_{i}^{*}-t_i $. Let the distance travelled when $ p_{i}(t)>0 $ is $ L_{i}^{*} $, we have
\begin{IEEEeqnarray}{rCl}
	L_{i}^{*}(x_i,a_i)=\int_{0}^{T_{i}^{*}}(x_{i}+a_{i}t)\,dt. \nonumber
\end{IEEEeqnarray}
Following the state transition function \eqref{state}, there is
\begin{IEEEeqnarray}{rCl}
	v_{i}(t^{*})=x_i+a_{i}T_{i}^{*}=\sqrt{x_{i}^{2}+2a_{i}L_{i}^{*}}. \nonumber
\end{IEEEeqnarray}
Substituting the above into \eqref{output3}, there is
\begin{IEEEeqnarray}{rCl}
	G_{i}^{+}(x_i,a_i) &=& \frac{L_{i}^{*}(x_i,a_i)}{\beta^{+}}\left(ma_i+\beta_{i}^{(0)}\right) +\frac{L_{i}^{*}(x_i,a_i)\beta^{\rm air}}{\beta^{+}}\left(x_{i}^{2}+a_{i}L_{i}^{*}(x_i,a_i)\right) \nonumber \\
	&=& \tilde{\gamma}_{i}^{(0)}(x_i,a_i)+\tilde{\gamma}_{i}^{(1)}(x_i,a_i)a_{i}+\tilde{\gamma}_{i}^{(2)}(x_i,a_i)x_{i}^{2}, \nonumber
\end{IEEEeqnarray}
where $ \tilde{\gamma}_{i}^{(0)}(x_i,a_i)=\frac{L_{i}^{*}\beta_{i}^{(0)}}{\beta^+} $, $ \tilde{\gamma}_{i}^{(1)}(x_i,a_i)=\frac{L_{i}^{*}}{\beta^{+}}(m+\beta^{\rm air}L_{i}^{*}) $, $ \tilde{\gamma}_{i}^{(2)}(x_i,a_i)=\frac{L_{i}^{*}\beta^{\rm air}}{\beta^{+}} $.

Here it is emphasized that the coefficients $ \tilde{\gamma}_{i}^{(0)} $, $ \tilde{\gamma}_{i}^{(1)} $, $ \tilde{\gamma}_{i}^{(2)} $, depend on the vehicle state $ (x_i, a_i) $.

To compute $ G_{i}^{-}(x_i,a_i) $, there is
\begin{IEEEeqnarray}{rCl}
	G_{i}^{-}(x_i,a_i) &=& \int_{t_{i}^{*}}^{t_{i+1}}p_{i}(t)\,dt=\frac{ma_{i}+\beta_{i}^{(0)}}{\beta^-}\int_{0}^{\bar{T}_{i}^{*}}(x_{i}^{*}+a_{i}t)\,dt + \frac{\beta^{\rm air}}{\beta^{-}}\int_{0}^{\bar{T}_{i}^{*}}(x_{i}^{*}+a_{i}t)^{3}\,dt, \nonumber
\end{IEEEeqnarray}
where $ \bar{T}_{i}^{*}=T_{i}-T_{i}^{*} $ and $ x_{i}^{*}=v_{i}(t^*) $. The distance traveled when $ p_{i}(t) < 0 $ is
\begin{IEEEeqnarray}{rCl}
	\bar{L}_{i}^{*}(x_i,a_i)=1-L_{i}^{*}(x_i,a_i)=\int_{0}^{\bar{T}_{i}^{*}}(x_{i}^{*}+a_{i}t) \,dt. \nonumber
\end{IEEEeqnarray}
Again, there is
\begin{IEEEeqnarray}{rCl}
	x_{i+1}=x_{i}^{*}+a_{i}\bar{T}_{i}^{*}=\sqrt{(x_{i}^{*})^2+2a_{i}\bar{L}_{i}^{*}}. \nonumber
\end{IEEEeqnarray}
Therefore, there is
\begin{IEEEeqnarray}{rCl}
	G_{i}^{-}(x_i,a_i) &=& \frac{\bar{L_{i}^{*}}}{\beta^{-}}\left(ma_{i}+\beta_{i}^{(0)}\right) + \frac{\bar{L_{i}^{*}}\beta^{\rm air}}{\beta^{-}}\left[(x_{i}^{*})^{2}+a_{i}\bar{L}_{i}^{*}\right] \nonumber\\
	&=& \frac{\bar{L_{i}^{*}}}{\beta^{-}}\left(ma_{i}+\beta_{i}^{(0)}\right) + \frac{\bar{L_{i}^{*}}\beta^{\rm air}}{\beta^{-}}\left[x_{i}^{2}+a_{i}\left(L_{i}^{*}+1\right)\right], \nonumber
\end{IEEEeqnarray}
where the following fact was used
\begin{IEEEeqnarray}{rCl}
	x_{i}^{*}=x_{i}+a_{i}T_{i}^{*}=\sqrt{x_{i}^{2}+2a_{i}L_{i}^{*}}, \; L_{i}^{*}+\bar{L}_{i}^{*}=1. \nonumber
\end{IEEEeqnarray}

Now, there is
\begin{IEEEeqnarray}{rCl}
	G_{i}^{-}(x_i,a_i)=\bar{\gamma}_{i}^{(0)}(x_i,a_i)+\bar{\gamma}_{i}^{(1)}(x_i,a_i)a_{i}+\bar{\gamma}_{i}^{(2)}(x_i,a_i)x_{i}^{2}, \nonumber
\end{IEEEeqnarray}
where $ \bar{\gamma}_{i}^{(0)}(x_i,a_i)=\frac{\bar{L}_{i}^{*}\beta_{i}^{(0)}}{\beta^{-}} $, $ \bar{\gamma}_{i}^{(1)}(x_i,a_i)=\frac{\bar{L}_{i}^{*}}{\beta^{-}}\left[m+\beta^{\rm air}\left(L_{i}^{*}+1\right)\right] $,\; $ \bar{\gamma}_{i}^{(2)}(x_i,a_i)=\frac{\bar{L}_{i}^{*}\beta^{\rm air}}{\beta^{-}} $.

Combing the two cases, there is
\begin{IEEEeqnarray}{rCl}\label{output31}
	G_{i} &=& G_{i}^{+}(x_i,a_i)+G_{i}^{-}(x_i,a_i) \nonumber\\
	&=& \gamma_{i}^{(0)}(x_i,a_i)+\gamma_{i}^{(1)}(x_i,a_i)a_{i}+\gamma_{i}^{(2)}(x_i,a_i)x_{i}^{2},
\end{IEEEeqnarray}
where $ \gamma_{i}(x_i,a_i)=\tilde{\gamma}_{i}(x_i,a_i)+\bar{\gamma}_{i}(x_i,a_i) $.

\textbf{Case IV:} The sign of $ F_{i}^{\rm track} $ changes from negative to positive, \emph{i.e.}, there exists $ t_{i}^{*} $ such that
\begin{IEEEeqnarray}{rCl}
	\begin{cases}
		F_{i}^{\rm track}(t) \leq 0, \; t_{i} \leq t \leq t_{i}^{*} \nonumber\\
		F_{i}^{\rm track}(t) \geq 0. \; t_{i}^{*} \leq t \leq t_{i+1} \nonumber
	\end{cases}
\end{IEEEeqnarray}

The derivation in this case follows the same as in \textbf{Case III} and thus shares the same expression for $ G_i(x_i,a_i) $ as \eqref{output31} with however different values of $ \gamma_i^{(0)} $, $ \gamma_i^{(1)} $ and $ \gamma_i^{(2)} $.

Therefore, the sign of track force within segment $ i $ is determined by the vector $ (x_i,a_i ) $, and the expressions of vector parameter $ \gamma_{i} $ under different cases are listed in Table \ref{gammaval}.
\begin{table}[h]
	\begin{threeparttable}	
	    \caption{Value of vector parameter $ \gamma_{i} $ under different cases}
	    \label{gammaval}
	    \setlength\tabcolsep{0pt}
	    \begin{tabular*}{\columnwidth}{@{\extracolsep{\fill}}ccccc}
	    	\toprule
	    	$ \gamma_i $ & Case I & Case II & Case III & Case IV \\
	    	\midrule
	    	$ \gamma_i^{(0)}$ & $ \frac{\beta_i^{(0)}}{\beta^+} $ & $\frac{\beta_i^{(0)}}{\beta^-}$ & $ \frac{L_i^* \beta_i^{(0)}}{\beta^+}+ \frac{\bar{L}_i^* \beta_i^{(0)}}{\beta^-}$ & $ \frac{L_i^* \beta_i^{(0)}}{\beta^-}+ \frac{\bar{L}_i^* \beta_i^{(0)}}{\beta^+}$ \\
	    	$ \gamma_i^{(1)} $ & $ \frac{m+\beta^{\rm air}}{\beta^+} $ & $ \frac{m+\beta^{\rm air}}{\beta^-} $ & $\begin{array}{l}
	    		 \frac{L_i^*}{\beta^+}(m+\beta^{\rm air} L_i^*)+\frac{\bar{L}_i^*}{\beta^-} \\ \left[m+\beta^{\rm air} \left(L_i^* + 1\right)\right]  \end{array}$ & $\begin{array}{l}
	    		\frac{L_i^*}{\beta^-}(m+\beta^{\rm air} L_i^*)+\frac{\bar{L}_i^*}{\beta^+} \\ \left[m+\beta^{\rm air} \left(L_i^* + 1\right)\right] \end{array}$ \\
    		$ \gamma_i^{(2)} $ & $ \frac{\beta^{\rm air}}{\beta^+} $ & $ \frac{\beta^{\rm air}}{\beta^-} $ & $ \frac{L_i^* \beta^{\rm air}}{\beta^+}+\frac{\bar{L}_i^* \beta^{\rm air}}{\beta^-} $ & $ \frac{L_i^* \beta^{\rm air}}{\beta^-}+\frac{\bar{L}_i^* \beta^{\rm air}}{\beta^+} $ \\
	    	\bottomrule
	    \end{tabular*}
	\end{threeparttable}
\end{table}


\section{Derivation of upper-speed limit} \label{app:traff}
A safe headway between the truck and the preceding vehicle must be maintained all the time to ensure the driving safety. That is, the maximum permissible speed of the truck in the next segment must be limited according to the speed of the preceding vehicle. The headway is defined as the time that elapses between the arrival of the leading vehicle and the following vehicle at the designated test point. Let’s use $ h_{\tau} $ to indicate the headway, and thus there is
\begin{IEEEeqnarray}{rCl}
	h_{\tau}=\frac{d}{v_f}, \nonumber
\end{IEEEeqnarray}
where $ d $ is the spacing between the two vehicles and $ v_f $ is the velocity of the following vehicle. To ensure the driving safety of the truck through segment $ i $, there is
\begin{IEEEeqnarray}{rCl}\label{safety}
	\begin{cases}
		\frac{d_i+T_i v_p-l }{x_{i+1}} \geq h_{\tau} \\
		T_i=\frac{2l}{x_i+x_{i+1}},
	\end{cases}
\end{IEEEeqnarray}
where $ d_i $ is the spacing from the preceding vehicle at the beginning of segment $ i $, $ l $ is the parameterized length of each road segment, and $ v_p $ is the velocity of the preceding vehicle and is assumed constant. Based on \eqref{safety} , then there is
\begin{IEEEeqnarray}{rCl}
	\begin{cases}
		x_{i+1} \leq \frac{L_i-h_{\tau}x_i+\sqrt{(h_{\tau}x_i-L_i)^2+4h_{\tau(L_i x_i+2lv_p)}}}{2h_{\tau}}=v_{\rm max}^{\rm hw} \\
		x_{i+1} \geq 0 \geq \frac{L_i-h_{\tau}x_i-\sqrt{(h_{\tau}x_i-L_i)^2+4h_{\tau(L_i x_i+2lv_p)}}}{2h_{\tau}},
	\end{cases}
\end{IEEEeqnarray}
where $ L_i=d_i-l, i=1,2,\cdots,N $. Therefore, the maximum permissible value of $ x_{i+1} $ during optimization is $ \bar{x}_{i+1}={\rm min}(v_{\rm max}, v_{\rm max}^{\rm hw}) $, where $ v_{\rm max} $ is the legally maximum permissible speed in EU.

\section*{Acknowledgment}
The authors would like to thank Dr. Hellström in Ford Motor Company for kindly sharing the road altitude data between the cities of Södertälj and Norrköping in Sweden, and the extensive discussions with engineers in Scania AB. This research is partly supported by Chalmers AoA Transport and EU JPI project SMUrTS.

\ifCLASSOPTIONcaptionsoff
  \newpage
\fi



%
\bibliographystyle{ieeetr.bst}
\bibliography{References.bib}

%






\end{document}